# Deterministically fabricated quantum dot single-photon source emitting indistinguishable photons in the telecom O-band

N. Srocka[1], P. Mrowiński[1,2], J. Große[1], M. von Helversen[1], T. Heindel[1], S. Rodt[1], S. Reitzenstein[1]

[1]*Institut für Festkörperphysik, Technische Universität Berlin, Hardenbergstraße 36, 10623 Berlin, Germany*

[2]*Laboratory for Optical Spectroscopy of Nanostructures, Department of Experimental Physics, Wrocław University of Technology, Wybrzeże Wyspiańskiego 27, Wrocław, Poland*

**Abstract**

In this work we develop and study single-photon sources based on InGaAs quantum dots (QDs) emitting in the telecom O-band. The quantum devices are fabricated using in-situ electron beam lithography in combination with the thermocompression bonding to realize a backside gold mirror. Our structures are based on InGaAs/GaAs heterostructures, where the QD emission is redshifted towards the telecom O-band at 1.3 μm via a strain reducing layer. QDs pre-selected by cathodoluminescence mapping are embedded into mesa structures with a back-side gold mirror for enhanced photon-extraction efficiency. Photon-autocorrelation measurements under pulsed non-resonant wetting-layer excitation are performed at temperatures up to 40 K showing pure single-photon emission which makes the devices compatible with stand-alone operation using Stirling cryocoolers. Using pulsed p-shell excitation we realize single-photon emission with high multi-photon suppression of $g^{(2)}(0) = 0.027 \pm 0.005$, post-selected two-photon interference of about (96 ± 10) % and an associated coherence time of (212 ± 25) ps. Moreover, the structures show an extraction efficiency of ~5 %, which compares well with values expected from numeric simulations of this photonic structure. Further improvements on our devices will enable implementations of quantum communication via optical fibers.

Single photons, often referred to as flying qubits, are key resource in the field of photonic quantum technology and enable for instance the distribution of quantum information over large distances.[1,2] Moving beyond simple point-to-point quantum key distribution protocols, such as BB84,[3] the quantum nature of single photons in terms of high photon indistinguishability becomes important. In fact, high photon indistinguishability is required in large-scale quantum networks based e.g. on entanglement distribution via Bell state measurements. Moreover, single photon emitters have to comply with existing optical fiber infrastructure for long-haul communication in the telecommunication O-band at 1.3 µm or in the C-band at 1.55 µm.[4–14] Therefore, besides a robust device concept with excellent quantum properties, the spectral matching to one of the two telecom bands is needed to enable fiber-based quantum communication of large distances. Despite significant progress,[6–9,13–20] it is still a great challenge to fulfill these stringent requirements with on-demand quantum dot (QD) based single-photon sources.

In this work, we address the above-mentioned requirements within a deterministic device technology. We transfer and optimize technology concepts, which allowed for high-quality emission from semiconductor QDs in the spectral range below 1 µm, to deterministically realize bright QD single-photon sources (SPSs) emitting in the telecom O-band. The sources are based on self-assembled InGaAs/GaAs QDs grown by metal organic chemical vapor deposition (MOCVD), where a strain-reducing layer (SRL) is applied to shift the emission wavelength to the telecom O-band.[10,21–23] Using in-situ electron beam lithography (EBL) such QDs are deterministically integrated into nanophotonic structures. Here, low-temperature cathodoluminescence (CL) spectroscopy is used to pre-select suitable QDs based on their emission wavelengths and brightness, before micromesas allowing for enhanced photon-extraction efficiency are patterned with 30-40 nm alignment accuracy[24] by in-situ EBL in the same system. The realized QD-micromesas with a backside gold mirror are designed to enable broadband enhancement of photon-extraction efficiency. This way, we realize high-quality telecom O-band SPSs with strong suppression of multi-photon emission events to 40 K. Moreover, by means of Hong-Ou-Mandel experiments we determine a two-photon interference (TPI) visibility of 96% under temporal post selection and of 12% without post-selection.

We developed an advanced flip-chip based processing concept for telecom SPSs in which the single-QD device is fabricated within three main processing steps. First, a GaAs semiconductor heterostructure is grown by MOCVD including a single layer of self-assembled

InGaAs QDs. Then, a thin membrane of this structure is transferred on a gold mirror by thermocompression bonding and wet-chemical etching, before single QDs inside this membrane are deterministically integrated into mesa structures by means of in-situ EBL.

The epitaxial growth starts with a 200 nm thick GaAs buffer on top of an n-doped GaAs (100) substrate to enable a high-quality epitaxial surface. Then, two $Al_{0.90}Ga_{0.10}As$ layers of 1 µm and 100 nm thickness are deposited. These two layers are separated by a 2 µm-thick GaAs spacer layer which allows for a well-controlled two-step wet-chemical etching procedure to remove the initial substrate. All of these layers act as etch-stop and sacrificial layers and will be removed during the flip-chip post-growth processing.[25,26] The growth resumes with 637 nm GaAs, 1.5 monolayers of $In_{0.7}Ga_{0.3}As$ followed by an 0.5 monolayer GaAs flush to realize the QD layer, 5.5 nm InGaAs, whereby the In-content is linearly decreased from 30 to 10 %, forming the SRL, and a final GaAs capping layer of 242 nm complete the layer structure. The described layer design is depicted in Fig. 1(a).

To enable thermocompression gold bonding a 250 nm thick gold layer is deposited on the as-grown heterostructure sample as well as on a bare piece of GaAs (100) substrate acting as host substrate. The sample is bonded to the host substrate (gold facing gold) by applying a pressure of 6 MPa and a temperature of approximately 600 K for four hours. The subsequent wet-chemical etching is performed in two steps. First, the 300 µm thick GaAs substrate is lifted off with a fast etching solution ($H_2O_2/NH_3$, 10:1), where the etching stops at the first $Al_{0.97}Ga_{0.03}As$ layer. After an HCl dip removing this etch-stop layer, the following GaAs layer is removed by a different etchant (citric acid/$H_2O_2$, 4:1). Here, the less aggressive etchant was chosen in favor of an improved surface roughness over a high etch rate. A final HCL dip leaves only a thin GaAs membrane of 885 nm, including the active QD layer, being gold bonded to the host GaAs substrate. On the cleaned sample 250 nm of AR-P 6200 electron-beam resist are spin coated in preparation for the lithography step.

In the subsequent main processing step, single pre-selected QDs are integrated deterministically into micromesas to enhance their photon extraction efficiency.[17,27,28] QD pre-selection is performed by CL mapping over (20 × 20) µm² sample areas at 10 K. Immediate after, mesa patterns are written into the resist at the chosen QD positions by EBL in the same system at 10 K. During the subsequent resist development, the mapping area is cleared and only EBL patterned areas remain, acting as masks in the subsequent inductively-coupled-plasma reactive-ion etching of QD mesas. We refer to Refs.[24,29] for further details of the in-situ EBL technique. Fig. 1(b) shows an optical microscopy image and panel (c) presents a CL map of deterministic QD mesa structures. The CL map in Fig. 1(c) was taken with the same settings as for the pre-selection step, clearly showing the fabrication of deterministic QD-micromesas.

The optical and quantum optical characterization of the QD mesas is performed by means of micro-photoluminescence (µPL), µPL-excitation (µPLE) and photon-correlation spectroscopy, respectively. The sample is mounted in a He-flow cryostat with temperature control in the range of 10 K to 40 K. Excitation is provided by a CW laser and tunable pulsed lasers focused on the sample by a microscope objective (numerical aperture (NA) = 0.4): A continuous-wave diode laser (785 nm) and a tunable laser providing ps-pulses at a repetition rate of 80 MHz. The photoluminescence signal is collected with the same objective, spectrally resolved in a grating spectrometer (spectral resolution ~ 0.05 nm) and either detected with a liquid nitrogen cooled InGaAs-array detector or coupled into a fiber-based Hanbury-Brown and Twiss (HBT) or a Hong-Ou-Mandel (HOM) setup. For HBT and HOM measurements the photons are detected by two superconducting nanowire single photon detectors (SNSPDs), each with a temporal resolution of ~50 ps and a detection efficiency of about 80% at 1310 nm.

First, temperature-dependent µPL spectra and photon autocorrelation histograms were recorded for the selected deterministic photonic microstructure marked in Fig. 1(c). Corresponding µPL spectra are presented in Fig. 2(a) and (b), respectively, for temperatures from 10 K to 40 K. Three excitonic states – exciton (X), biexciton (XX) and charged exciton (CX) – of the single-QD mesa are identified by excitation-power- and polarization-dependent µPL studies (see supplementary material – Fig. S1). A temperature-induced red-shift is observed together with a notably stable single-QD emission up to 40 K, reflecting a rather deep carrier confinement in this type of QDs. The temperature stability is an important aspect regarding the development of stand-alone SPS operated by Stirling cryocoolers with a base temperature between 30 K and 40 K (depending on the manufacturer).[30–32] The brightest CX emission line is further examined in the HBT setup to investigate the quantum nature of emission. Here, we obtain high multi-photon suppression of $g^{(2)}(0)_{fit} = 0.076 \pm 0.015$ at 10 K and $g^{(2)}(0)_{fit} = 0.114 \pm 0.022$ at 40 K under pulsed non-resonant wetting-layer excitation (974 nm), as shown in Fig. 2(b). The fit presented in Fig. 2(b) is based on a mono-exponential decay convoluted with the instrument response function (IRF) of the detection system. The fit yields a background signal of about 0.05 in $g^{(2)}(\tau)$ due to uncorrelated background emission.

Noteworthy, the lifetime of the CX state varies in the investigated temperature range between ~1.6 ns and ~1.9 ns with no clear trend, although state refilling processes from higher levels[29] influencing the occupation per excitation pulse of the CX state are seen for higher temperatures (see supplementary material – Fig. S2).

Next, we examined the source brightness in terms of the photon-extraction efficiency $\eta_{\text{ext}}$. The QD was excited by pulsed non-resonant (~860 nm) excitation at saturation of the CX intensity. Under those conditions, we estimate a lower bound of $\eta_{\text{ext}}$ assuming that the internal QD efficiency is close to one (i.e. non-radiative rate is negligible). The emitted photons were detected by SNSPDs with an overall setup efficiency of $\eta_{\text{setup}} \sim 5.4\%$. For the given laser pulse repetition rate of 80 MHz we obtained 172.000 counts per seconds (cps) for the sum of X and CX emission yielding an extraction efficiency of $\eta_{\text{ext}} = \frac{cps}{f\,\eta_{\text{setup}}} = 5.0\%$. This lower bound value compares well with the expected extraction efficiency of 9.3% deduced from finite-element method simulations for this particular mesa (see supplementary material).

High photon indistinguishability is an important requirement for many applications of SPSs in quantum technology. The TPI effect is measured for photons emitted by the QD-micromesa under study using a fiber-based HOM setup consisting of an unbalanced Mach–Zehnder interferometer (MZI) on the detection side and a complementary MZI on the excitation side. Here, the 4 ns relative optical path delay of the MZI in the detection path is compensated by the temporal separation of excitation pulses induced by the MZI on the excitation path.

Before performing HOM experiments we established pulsed quasi-resonant p-shell excitation for the QD-micromesa under study to reduce charge noise and detrimental dephasing processes related to carrier relaxation from higher levels, presumably taking place mainly in the SRL. The s-p energy splitting is determined via μPLE experiments. As observed in Fig. 3(a) the excitation laser is in resonance with the QD's p-shell at 1.0225 eV. The corresponding s-p splitting amounts to $\Delta E_{\text{s-p}} = 65.7$ meV (see also supplementary material – Fig. S3) and is in good agreement with values reported in Ref.[4,34] When exciting quasi-resonantly at an excitation power required to achieve ~70 % of the maximum CX intensity, we achieved a further improvement of the $g^{(2)}(0)$ value. In Fig. 3(b) we fit the correlation data using again a mono-exponential decay convoluted with the IRF of the setup and with an uncorrelation background of 0.023, which yields $g^{(2)}(0)_{\text{fit}} = 0.027 \pm 0.005$ for the telecom O-band QD-micromesa. In addition, we observe noticeable blinking in the side maxima up to time delays τ of about 40 ns, which we attribute to the memory effect for the subsequent pulses due to local charge fluctuations.[29] Within ±3 ns around zero delay a slight recapture process is also indicated by a minimum observed at τ = 0. Noteworthy, the achieved $g^{(2)}(0)$ value of 0.027 ± 0.005 is similar to the lowest value ($g^{(2)}(0)$ = 0.03) observed previously for InGaAs/GaAs QDs emitting in the telecom O-band[4] and proofs that our advanced device, processed in multiple steps, maintains a single-photon emission with high suppression of multi-photon events. Further improvements are expected for two-photon resonant excitation schemes.[30,31]

In Fig. 3(c) and (d) we present TPI histograms obtained for the CX line under p-shell excitation. As we expect an overlap of the side peaks of subsequent laser pulses, it is convenient to measure both the interference histogram for co- and cross-polarized photons and then to quantify the central peak suppression by direct comparison.[38] According to the fitting function based on a series of Lorentzian peak functions, the histogram in Fig. 3(d) shows a single peak centered at τ = 0 with 2.693 coincidences for cross-polarized photons. In case of co-polarized photons in Fig. 3(c), we observe a suppression of coincidences down to 0.114. These numbers yield a post-selected TPI visibility $V = (g^2(0)_\perp - g^2(0)_\parallel)/g^2(0)_\perp$, of (96 ± 10)%. Taking into account the integrated peak areas we obtain a visibility of (12 ± 4)% without post-selection. Moreover, by using the post-selected $V$ as an input parameter for a description based on mono-exponential decay functions[8] convoluted with the IRF we obtain a coherence time $\tau_c$ = (212 ± 25) ps (see supplementary material for details). These results show that further optimization and research is required to improve the coherence time and the indistinguishability. Most probably, the experimental values achieved are limited by structural imperfections and charge noise introduced by the SRL and HOM studies can be used as sensitive tool in the technological optimization of QDs emitting at wavelengths around 1.3 μm. Noteworthy, our results for QD-micromesas feature significantly higher post-selected indistinguishability and coherence time than 67% and 150 ps reported previously for a non-resonantly driven 1.3 μm QD coupled to a PhC nanocavity.[8] Moreover, our photonic structures support broadband enhancement of photon extraction and are very robust which makes them suitable for development of stand-alone SPSs.

In summary, we have shown a deterministically fabricated single-photon device emitting in the telecom O-band. The single-QD micromesa with a backside gold-mirror features single-photon emission stable up to 40 K, which makes it compatible with Stirling cryocoolers for stand-alone operation. The device shows strong multi-photon suppression associated with $g^{(2)}(0)$ as low as 0.027 under pulsed p-shell excitation. The generated photons show indistinguishability of 96% and 12% with and without post-selection, respectively. The broadband design of the microstructure facilitates a photon-extraction efficiency of ~5%. This value compares well with the simulated extraction efficiency of about 9%. The developed flip-chip based telecom-SPS device technology is compatible with spectral strain tuning when combined with a lower piezo-actuator[39] and with circular Bragg resonator SPSs which promises significantly higher photon-extraction efficiency in the future.[40] In summary, we developed and deterministically fabricated a SPS emitting in the telecom O-band, which meets important requirements for single-photon sources for application in fiber-based quantum communication. Combining these requirements, especially for the emission in the O-band, is a significant step towards versatile high-performance single-photon sources.

**Supplementary Material**

See the supplementary material for information on the identification of excitonic complexes in quantum dots, the mission dynamics and spontaneous emission lifetime of quantum dots, micro-photoluminescence excitation spectroscopy, and on optical studies under quasi-resonant p-shell excitation. Moreover, the supplementary material includes details on the numerical optimization of QD-micromesas and the evaluation of Hong-Ou-Mandel experiments.

**ACKNOWLEDGEMENTS**

The research leading to these results has received funding from the German Research Foundation through CRC 787 "Semiconductor Nanophotonics: Materials, Models, Devices", the Volkswagen Foundation via the project NeuroQNet, and the FI-SEQUR project jointly financed by the European Regional Development Fund (EFRE) of the European Union in the framework of the programme to promote research, innovation and technologies (Pro FIT). T. H. gratefully acknowledges financial support of the German Federal Ministry of Education and Research (BMBF) via the project 'QuSecure' (Grant No. 13N14876) within the funding program Photonic Research Germany. P.M. gratefully acknowledges financial support from the Polish Ministry of Science and Higher Education within "Mobilność Plus – V edycja" program.

DATA AVAILABILITY

The data that support the findings of this study are available from the corresponding author upon reasonable request.

Figures

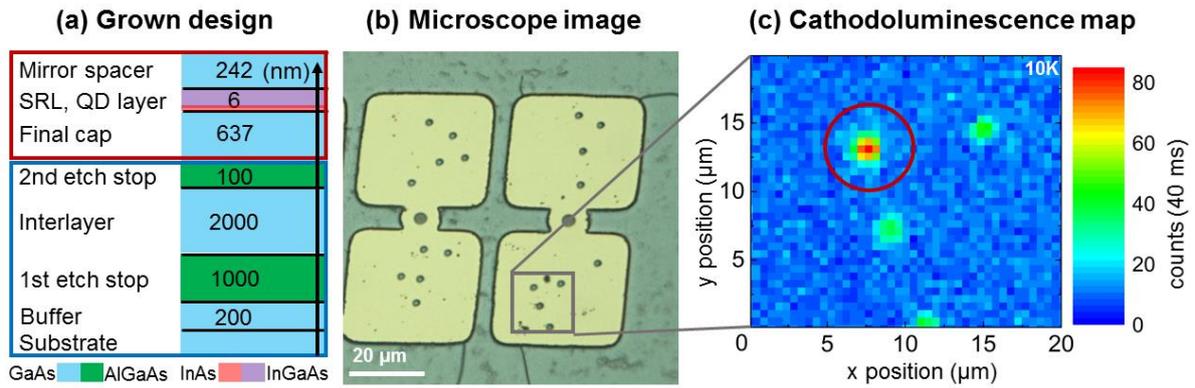

FIG. 1. (a) Schematic epitaxial layer design. Layers in the blue box are etch-stop / sacrificial layers removed during the post-growth processing. After a flip-chip process the layers in the red box are bonded up-side-down to a gold mirror. For this reason, the given layer names refer to their final device functionality. (b) Optical microscopy image of the final device. Four to five mesas are processed per mapping field. (c) Low temperature CL map recorded after full processing (integration time/pixel: 40 ms, spectral range depicted: target wavelength ± 1 nm). The red circle marks the QD-micromesa investigated in the following.

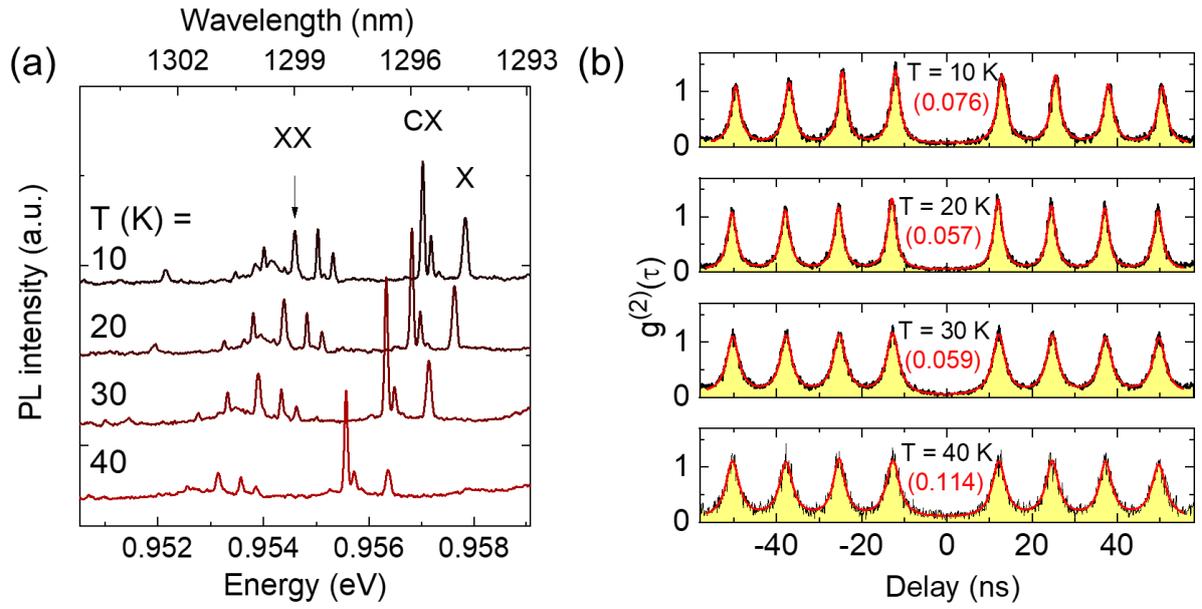

FIG. 2. (a) Temperature-dependent µPL spectra of exciton (X, CX) and biexciton (XX) complexes recorded under non-resonant CW excitation at 785 nm. (b) Corresponding $g^{(2)}(\tau)$ histograms and extracted $g^{(2)}(0)$ values (in red) of the CX line.

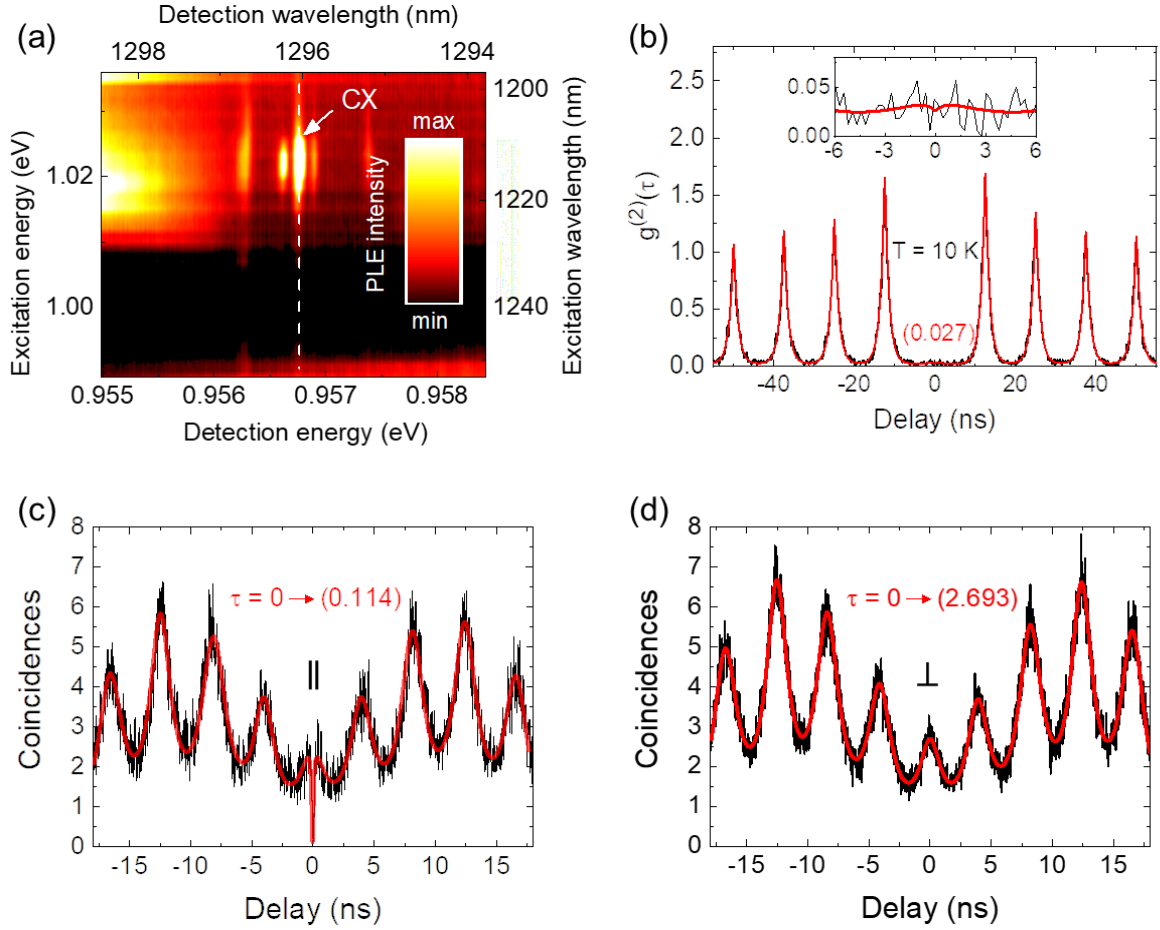

FIG. 3. (a) μPLE data measured on a deterministic QD microstructure demonstrating spectral resonance with the p-shell at 1.022 eV for excitonic states, i.e. CX emitting at 0.956 eV as indicated by the dashed line. (b) Corresponding pulsed second-order photon autocorrelation measurement at 10 K showing clean single-photon emission with $g^{(2)}(0) = 0.027 \pm 0.005$ according to the fitting function (red) superimposed to the raw histogram data. (c) and (d) two-photon interference histograms measured for the CX emission under pulsed p-shell excitation for co- and cross-polarized configuration, respectively. The HOM-effect is proven by the highly reduced coincidences in (c) compared to (d). The red trace corresponds to a fit function used to evaluate the two-photon interference visibility (see supplementary material for details).

Supplementary materials

# Deterministically fabricated quantum dot single-photon source emitting indistinguishable photons in the telecom O-band

N. Srocka[1], P. Mrowiński[1,2], J. Große[1], M. von Helversen[1], T. Heindel[1], S. Rodt[1], S. Reitzenstein[1]

[1]*Institut für Festkörperphysik, Technische Universität Berlin, Hardenbergstraße 36, D-10623 Berlin, Germany*
[2]*Laboratory for Optical Spectroscopy of Nanostructures, Department of Experimental Physics, Wrocław University of Technology, Wybrzeże Wyspiańskiego 27, Wrocław, Poland*

**1. Identification of excitonic complexes in quantum dots**

Identification of basic excitonic states in quantum dots (QDs) is performed with excitation power dependence and polarization series in linear basis, as shown in Fig. S1. In case of non-resonant excitation at 785 nm we found distinct emission lines in the low excitation power regime, which are attributed to the neutral (X) and the charged exciton (CX). With increasing excitation power $P$, the emission intensity $I$ increases proportional to $\sim P^S$, where $S$ is the slope parameter. For the three lines, already visible at the lowest excitation power $P_0$, $S \cong 1$ holds. At higher excitation power regime, we observe additional emission lines at higher wavelengths (lower energy) most probably coming from higher excitonic states of the same QD. For the line at 0.9546 eV the well-known characteristic slope change for biexcitons (XXs), could be proven with $S \cong 2$. In addition, the polarization series in linear basis exhibits a fine structure splitting (FSS) of 40 μeV for X and XX. As expected, the CX line exhibits no FSS.

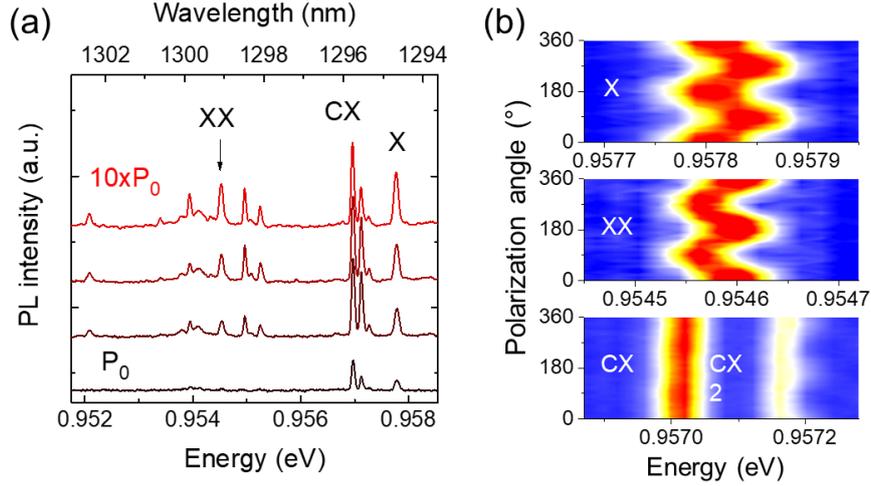

FIG. S1: µPL spectra of the deterministic QD-micromesa in dependence on excitation power (a) and linear polarization (b) (X - neutral exciton, XX – biexciton, CX - charged exciton)

## 2. Emission dynamics and spontaneous emission lifetime

We investigated the photoluminescence lifetime for the CX emission line of the QD-micromeas from 10 K to 40 K, as shown in Fig. S2. The pulsed laser was set to non-resonant wetting-layer excitation with the power sufficient to saturate the CX emission line. We observe state refilling processes from higher shells[1] as indicated by the peak maximum shift in the observed decay with increasing temperature, compare Fig. S2(a). Fits to the fast initial decays reveal emission lifetimes between 1.6 ns and 1.9 ns. These values are used as input data for the fitting of the presented autocorrelation measurements, see main text Fig. 2(b).

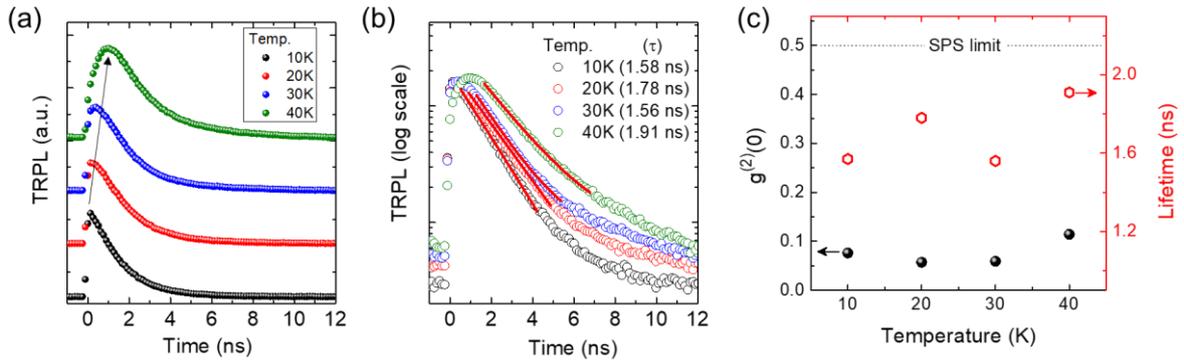

FIG. S2. Measurements of CX emission lifetime in dependence on temperature from 10 K to 40 K. In (a) the stack of the decay demonstrate the shift of the peak maximum, in (b) the fit to the experimental data is shown and in (c) the summary of the results both for $g^{(2)}(0)$ and the decay time is shown.

## 3. Micro-photoluminescence excitation spectroscopy

A micro-PL excitation (µPLE) experiment is performed to determine the energy difference between the QD's electronic s-shell ground state and the p-shell excited state, which might be excited by the detuned laser excitation. In Fig. S3(a) we show a µPLE map of the QD emission versus the excitation energy of the pump laser. We observe a pronounced maximum of the CX intensity at an excitation laser energy of 1.0225 eV, as shown in Fig. S3(b), which allows us to determine an s-p energy splitting of 65.7 ± 0.1 meV for this QD. The obtained results are used for an efficient CX excitation for autocorrelation and two-photon interference measurements.

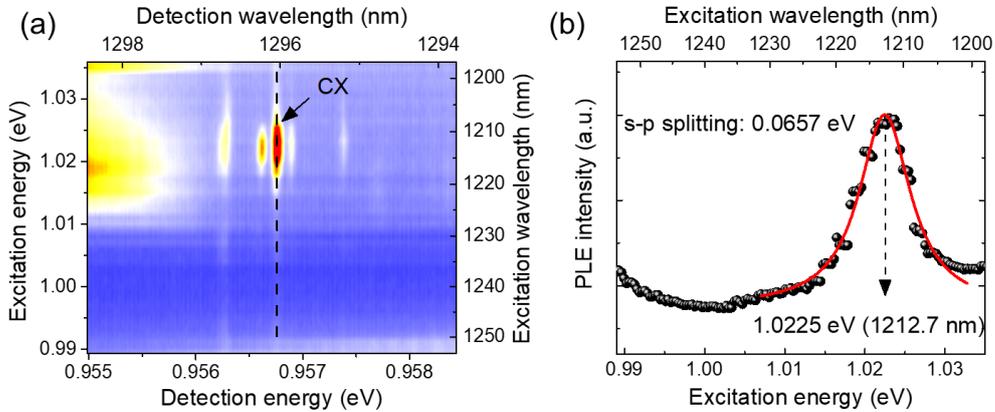

FIG. S3. (a) µPLE experiment performed for our QD-microlens showing the dependence of the emission intensity for excitonic states on the excitation energy of the laser. The analysis in (b) shows the Lorentzian fit of a µPLE intensity trace of the CX line that allows to determine the s-p energy splitting.

## 4. Optical studies under quasi-resonant p-shell excitation

We performed power dependent µPL measurements under p-shell excitation (exc. energy 1.0225 eV) on the QD-micromesa under investigation. The spectra are shown in Fig. S4(a). First, we determine the saturation of the emission intensity of the CX line and, second, we observe a red-shift of the emission with increasing excitation power. This spectral shift is most probably due to local heating of the sample, as the band-gap energy decreases with the temperature increase. A summary of the analysis is shown in Fig. S4 (b) where $P_{sat}$ and $P_{0.7sat}$ indicate the level of excitation power of the saturated CX intensity and 70% of the saturation, respectively.

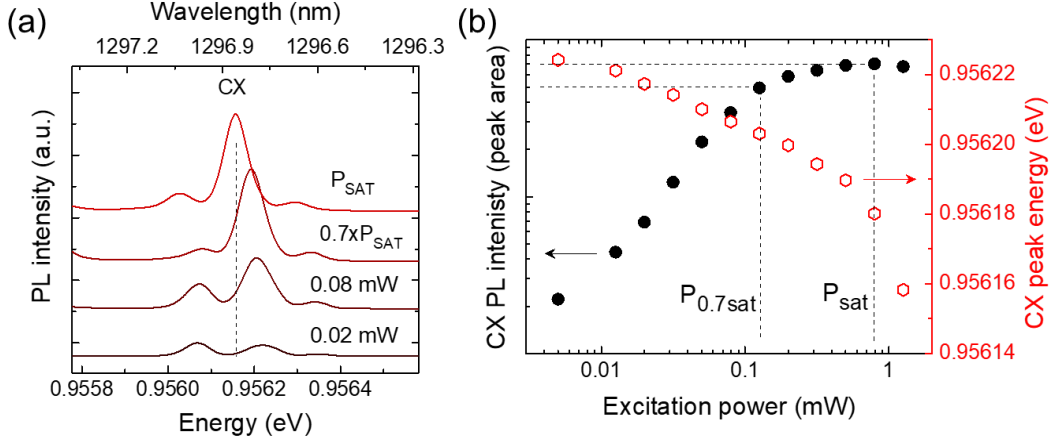

FIG. S4. (a) µPL spectra for p-shell excitation versus excitation power and (b) the emission intensity and peak energy analysis for the CX emission.

## 5. Numerical studies on the photon extraction efficiency of QD-micromesas

We performed a FEM based numerical analysis of the dipole emission from the photonic mesa structure with the commercially available software package JCMsuite from JCMwave[2]. The explicit mesa geometry is obtained by SEM imagining and is used as the numerical simulation input of a rotation symmetric device. The mesa with a height of 885 nm is trapezoidal with a base diameter of 1474 nm and tapers towards the top to 1220 nm in diameter. Fig. S5(a) shows the mesa sitting on a gold layer of about 500 nm thickness, with the enclosed dipole (QD) itself elevated to 242 nm above the gold layer. The dipole's emission wavelength, 1295 nm, as well as the refractive index of GaAs $n_{GaAs}$ = 3.406 are fixed parameters. In Fig. S5(b) the cross-section of an absolute electric field distribution (near-field) of the dipole emission is shown.

The extraction efficiency $\eta_{ext}$ is then determined by considering the calculated far-field. It is defined as $\eta_{ext} = \frac{P_{dip,NA}}{P_{dip}}$, where $P_{dip,NA}$ is the power of the field emitted by the dipole towards normal direction within the selected numerical aperture (NA), and $P_{dip}$ is the total dipole power. Within the analysis of the photon extraction efficiency of the mesa, we calculated that for NA = 0.4 and our structure's theoretical upper limit of the extraction efficiency is 9.3% (for NA = 0.6: 24.3% and for NA = 0.8: 42.8%), as visualized in Fig. S5(b). This compares well to our experimental findings, especially as the experimental value may be slightly reduced due to non-ideal quantum efficiency and an in-plane alignment mismatch of the QD position with respect to the mesa center.[3] Based on the developed device fabrication technology, significantly higher photon-extraction efficiencies should be possible in the future when using circular Bragg cavity structures[4].

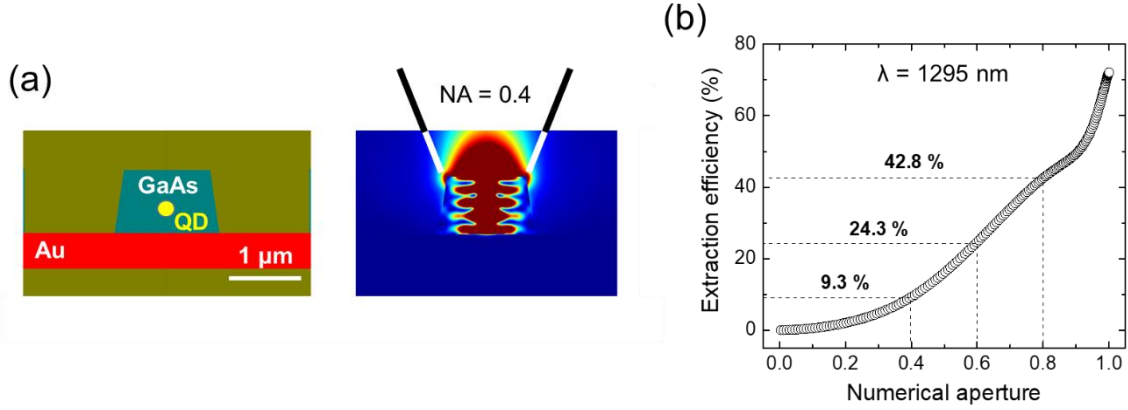

FIG. S5. (a) model of the photonic structure fabricated with the in-situ EBL deterministic technology and gold bonding (exact geometry is given in the text) together with the calculated near-field intensity of the dipole emission at 1295 nm. (b) The calculated extraction efficiency versus the numerical aperture.

6. **Hong-Ou-Mandel experiment - two-photon interference visibility**

The TPI effect is measured for photons emitted by the QD-micromesa under study using a fiber-based HOM setup consisting of an unbalanced Mach–Zehnder interferometer (MZI) on the detection side and a complementary MZI on the excitation side, both realizing the 4 ns relative optical path delay. Based on the 80 MHz repetition rate of the pulsed laser, we expect clusters of five peaks in the correlation histogram every 12.5 ns. Relative peak areas of the center cluster including contributions of neighboring clusters are proportional to 5:3:2:3:5 with respect to the time delay Δt = -8:-4:0:4:8 ns for distinguishable photons and to 5:3:0:3:5 for indistinguishable photons.[5] Therefore, for distinguishable photons we can expect a ratio of $A_0/A_1 = 2/3 = 0.667$ and $A_0/A_2 = 2/5 = 0.4$, where $A_{0,1,2}$ are integrated peak areas at the center (τ = 0 ns), ±4 ns and ±8 ns of the HOM histogram, respectively. Considering integrated peak areas, we can calculate the HOM visibility with

$$V = \frac{A_0^\perp - A_0^\parallel}{A_0^\perp} = \frac{\frac{2}{3}A_1^\parallel - A_0^\parallel}{\frac{2}{3}A_1^\parallel}$$

In order to find analytical expressions that reproduce the $g^{(2)}(\tau)^{HOM}_{\perp}$ and $g^{(2)}(\tau)^{HOM}_{\parallel}$ histograms we first use the following functions based on Lorentzian peak functions:

$$f_{\perp}(x) = \frac{2}{\pi}\left[A_0 \frac{w_0}{4(x-x_0)^2+w_0^2} + \sum_{i=1}^{4} A_i \frac{w_i}{4(x+x_i)^2+w_i^2} + \sum_{i=1}^{4} A'_i \frac{w_i}{4(x-x_i)^2+w_i^2}\right],$$

$$f_{\parallel}(x) = \frac{2}{\pi}\left[A_0 \frac{w_0}{4(x-x_0)^2+w_0^2} + A_{00} \frac{w_{00}}{4(x-x_0)^2+w_{00}^2} + \sum_{i=1}^{4} A_i \frac{w_i}{4(x+x_i)^2+w_i^2} + \sum_{i=1}^{4} A'_i \frac{w_i}{4(x-x_i)^2+w_i^2}\right]$$

Where $y_0$ is the backgound level, $A_{0-4}$ and $A'_{0-4}$ is the peak amplitude, $w_{0-4}$ is the width and $x_{0-4}$ is the peak position. Indices '0-4' are used to label peaks at delay of 0, ±4, ±8, ±12.5, ±16.5 ns, respectively. In addition $A_{00}$ and $w_{00}$ in $f_{\parallel}(x)$ defines the amplitude and width of the HOM dip at τ = 0 ns, and we expect for indistinguishable photons that $A_{00} < 0$ and $w_{00} << w_{0,1,2,3,4}$ in order to minimize fit residuals.

The fit is presented in Fig. 3c) and d) in the main text for parallel and orthogonal polarizations, respectively. Fit parameters concerning integrated areas of the peaks at -8:-4:0:4:8 ns for parallel polarization (∥) are:

$$A_{00,\parallel} = -0.74 \pm 0.15; A_{0,\parallel} = 6.77 \pm 0.46;$$
$$A_{1,\parallel} = 10.84 \pm 0.36; A'_{1,\parallel} = 9.66 \pm 0.33;$$
$$A_{2,\parallel} = 16.70 \pm 0.36; A'_{2,\parallel} = 17.90 \pm 0.31,$$

and for orthogonal polarization (⊥) they are:

$$A_{0,\perp} = 6.43 \pm 0.31;$$
$$A_{1,\perp} = 10.13 \pm 0.31; A'_{1,\perp} = 12.02 \pm 0.31;$$
$$A_{2,\perp} = 17.31 \pm 0.31; A'_{2,\perp} = 18.04 \pm 0.31.$$

We can verify the fits by checking $\frac{2A_0}{A_1+A'_1} = 0.661 \pm 0.048$, which is very similar to the expected value of $\frac{2}{3} = 0.667$, as well as $\frac{2A_0}{A_2+A'_2} = 0.391 \pm 0.027$, again similar to expected $\frac{2}{5} = 0.4$. The TPI visibility can be defined in several ways, using central peak only e.g.:

$$V = \frac{A_{0,\parallel}-(A_{0,\parallel}+A_{00,\parallel})}{A_{0,\parallel}} \times 100\% = (10.9 \pm 9.4)\%,$$

$$V = \frac{A_{0,\parallel} - (A_{0,\perp} + A_{00,\parallel})}{A_{0,\parallel}} \times 100\% = (16.0 \pm 7.7)\%,$$

$$V = \frac{A_{0,\perp} - (A_{0,\parallel} + A_{00,\parallel})}{A_{0,\perp}} \times 100\% = (6.2 \pm 8.8)\%,$$

$$V = \frac{A_{0,\perp} - (A_{0,\perp} + A_{00,\parallel})}{A_{0,\perp}} \times 100\% = (11.5 \pm 6.9)\%,$$

Or with the use of both central and first side peaks at ±4 ns, e.g.:

$$V = \frac{\frac{2}{3}A_{1,\parallel} - (A_{0,\parallel} + A_{00,\parallel})}{\frac{2}{3}A_{1,\parallel}} \times 100\% = (11.8 \pm 8.3)\%,$$

$$V = \frac{\frac{2}{3}A_{1,\perp} - (A_{0,\parallel} + A_{00,\parallel})}{\frac{2}{3}A_{1,\perp}} \times 100\% = (18.3 \pm 7.4)\%,$$

$$V = \frac{\frac{2}{3}A_{1,\parallel} - (A_{0,\perp} + A_{00,\parallel})}{\frac{2}{3}A_{1,\parallel}} \times 100\% = (16.7 \pm 6.5)\%,$$

$$V = \frac{\frac{2}{3}A_{1,\perp} - (A_{0,\perp} + A_{00,\parallel})}{\frac{2}{3}A_{1,\perp}} \times 100\% = (8.2 \pm 11.8)\%.$$

One can see, that depending on the approach, we obtain a distribution of the results, therefore we can estimated visibility using average of all cases, which results in $V_{ave} = (12.4 \pm 4.3)\%$. The uncertainity here is evaluated by using a standard deviation. The post selected visibility can be calculated according to fitting functions $f(x)$ as

$$V_{p\_s} = \frac{f_\perp(0) - f_\parallel(0)}{f_\perp(0)} \times 100\% = 96 \pm 10\%.$$

The width of the HOM dip is characterized by a coherence time $\tau_c$, while the overlapping central peak typical for distinguishable photons defined by $A_0$ and $w_0$ is characterized by radiative spontaneous emission lifetime $\tau_1$. In the ideal case, where all dephasing processes are neglegted, we expect $\frac{\tau_c}{2\tau_1} = 1$, however with pure dephasing characterized by $\tau_{deph}$, we should follow the relation of $\frac{1}{\tau_{deph}} = \frac{1}{\tau_c} - \frac{1}{2\tau_1}$ indicating that $\frac{\tau_c}{2\tau_1} < 1$. For indistinguishable photons we can evaluate the coherence time by using another fitting function based on mono-exponential decay convoluted with the timing response of the detection system, characterized by the jitter of ~50 ps. We use the following fitting function for the central and ±4 ns HOM peaks:

$$f(\tau) = \left[ y_0 + A_0 \cdot e^{-\frac{|\tau|}{\tau_1}} \left( 1 - V \cdot e^{-2\frac{|\tau|}{\tau_{deph}}} \right) + A_1 \cdot e^{-\frac{|\tau - d\tau|}{\tau_1}} + A'_1 \cdot e^{-\frac{|\tau + d\tau|}{\tau_1}} \right] \otimes G(\tau, \sigma_{res}),$$

where $d\tau \sim 4\ ns$ is the delay corresponding to the first side peak, V is the TPI visibility, and Gaussian function $G(\tau)$ is characterized by $\sigma_{res} = \frac{50\ ps}{2\sqrt{2\ln 2}}$. To evaluate of $\tau_c$ we fix $\tau_1 = 1.6\ ns$ according to lifetime measurements, as shown in Fig. S2, we set $V = V_{p\_s} = 0.96$ and $\frac{A_0}{A_1} = \frac{A_0}{A'_1} = \frac{2}{3}$, so the remaining fit parameters are $y_0$, $\tau_{deph}$. In this case, we obtain best fit for $\tau_{deph} = (224 \pm 20)\ ps$ and $y_0 \ll 0.01$, which gives $\tau_c = (212 \pm 25)\ ps$. The fit is shown in Fig. S6. It is worth noting here, that due to a slow decay of the side peak with its amplitude of $\frac{3}{2}A_0$ that can influence the proximity of the HOM dip, the exact fit to the minimum value of the as-measured coincidences, enabling the visibilty parameter $V$ to be free, requires $y_0 \lesssim 0$ and $\tau_1 \sim 2\ ns$, which results in $V \times 100\% \rightarrow 100\%$, however due to this unphysical set of parameters we are not able to determine its uncertainity.

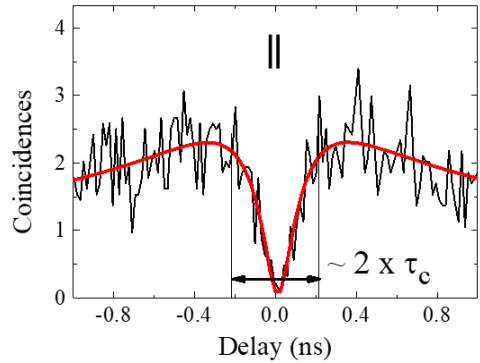

Fig S6. HOM histogram for parallel polarization of two photons showing a TPI dip resolved by mono-exponential decay functions convoluted with the instrument response function of the detection system, which allows to evaluate the coherence time.